\documentclass[twocolumn,prd,groupedaddress,nofootinbib,showpacs]
{revtex4}

\usepackage{latexsym} 
\usepackage{amssymb}  
\usepackage{graphicx} 
\usepackage{dcolumn}  
\usepackage{bm}       

\begin{document}

\title{Time-dependent cosmological constant in the
Jackiw-Teitelboim cosmology}

\author{M. Alves}
\email{msalves@if.ufrj.br}
\affiliation{Instituto de F\'{\i}sica\\
Universidade Federal do Rio de Janeiro, RJ 21945-970 -- Brazil}
\author{J. Barcelos-Neto}
\email{barcelos@if.ufrj.br}
\affiliation{Instituto de F\'{\i}sica\\
Universidade Federal do Rio de Janeiro, RJ 21945-970 -- Brazil}
\author{M. Novello}
\email{novello@cbpf.br}
\affiliation{Centro Brasileiro de Pesquisas F\'{\i}sicas,\\
Rua Dr.\ Xavier Sigaud 150, Urca 22290-180 Rio de Janeiro, RJ --
Brazil}
\author{J.M. Salim}
\email{jsalim@cbpf.br}
\affiliation{Centro Brasileiro de Pesquisas F\'{\i}sicas,\\
Rua Dr.\ Xavier Sigaud 150, Urca 22290-180 Rio de Janeiro, RJ --
Brazil}

\date{\today}

\begin{abstract}
We study the obtainment of a time-dependent cosmological constant at
$D=2$ in a model based on the Jackiw-Teitelboim cosmology. We show
that the cosmological term goes to zero asymptotically and can induce
a nonsingular behavior at the origin.
\end{abstract}


\maketitle

\section{Introduction}

The presence of cosmological constant is an interesting and intriguing
subject in many cosmological models~\cite{Weinberg}. If it actually
exists in the present stage of the universe, it has to be very small.
But the point is that it could have not been small since the
beginning. Consequently, realistic models to describe the evolution of
the universe should include a time-dependent of the cosmological {\it
constant}~\cite{Ozer}.

\medskip
In a previous paper~\cite{Novello1}, three of us have presented a
model where the origin of this time dependence was  related to a
possible quantum scenario of the initial evolution of the universe.
This was achieved by showing that the chiral gauge anomaly could be
conveniently adapted in order to generate a time-dependent
cosmological {\it constant}. In this way, the presence of this term
today would be a reminiscence of that initial quantum behavior. Later
on, this model was applied~\cite{Novello2} in a particular scenario
where the geometry was initially Bianchi-like (spatially homogeneous
but anisotropic)~\cite{Belinsky}. We have shown that the Einstein
equations in the presence of the cosmological term (as well as the
Maxwell one) leads to an asymptotic solution that is compatible with
the Friedmann universe~\cite{Weinberg,Novello3}.

\medskip
In these two papers, the time-dependence we have mentioned was
implicit into the possible quantum behavior of the initial evolution
of the universe, but we have not found this time dependence
explicitly. It could have been achieved if the equations of motion of
the theory were completely solved. However, since we are involved with
nonlinear equations, this is not an easy task.

\medskip
We address to this problem in the present paper. We try to circumvent
the algebraic difficulties by working in a gravitational theory in
spacetime dimension $D=2$, in the formulation given by Jackiw and
Teitelboim \cite{Jackiw}. We deal with two distinct situations in
order to implement the time cosmological constant in the model. The
first one is considered by following the same steps of reference
\cite{Novello2} adapted to $D=2$. We find that the only possible
solution is an actually cosmological constant. Even though there is no
time dependence for the cosmological term at $D=2$, the development of
this part will be useful to envisage the asymptotic solution of the
next case. Secondly, we use a Chern-Simons \cite{Deser} term at $D=2$
\cite{Barcelos} as the starting point to generate the cosmological
term. Even though the equations of motion we obtain are much simpler
than the four dimensional case, they are still very complicated to be
completely solved. However, with the help of the solution we have
obtained into the first case, we show that it is possible to have a
cosmological term which tends to zero as the time goes to infinity.
This is the behavior we would expect for a time-dependent cosmological
constant. We also consider its behavior closed to the origin, where
the presence of the cosmological {\it constant} permit us to have
both singular and nonsingular solutions, depending on the boundary
conditions we use.

\medskip
Our paper is organized as follow. In Sec. II we review the general
features of the model adapted to the two-dimensional case. The
applications mentioned above are displayed in Sects. III. and IV
respectively. We left Sec. V for some concluding remarks.

\section{Time-dependent cosmological constant in the Jackiw-Teitelboim
cosmology}
\renewcommand{\theequation}{2.\arabic{equation}}
\setcounter{equation}{0}

It is well-known that the Einstein-Hilbert action in $D=2$ is a
topological invariant quantity, in a sense that its corresponding
Lagrangian is a total derivative. So, there is no way to extract any
dynamics from it. The formalism due to Jackiw and Teitelboim
\cite{Jackiw} starts from a kind of two-dimensional version of the
Einstein equation (but with no action as origin), i.e.

\begin{equation}
R-\Lambda=8\pi\,G\,T
\label{2.1}
\end{equation}

\noindent
where $R$ is the Ricci scalar, $\Lambda$ is a cosmological constant,
$G$ is the gravitational constant and $T$ is the trace of the
energy-momentum tensor.

\medskip
It is opportune to mention that in the original Jackiw-Teiltelboim
model the right-hand side of (\ref{2.1}) was zero. The introduction of
the trace of the energy-momentum tensor into the equation was first
done by Mann et al.~\cite{Mann}.

\medskip
Let us now review how the time-dependent $\Lambda$ can be
generated~\cite{Novello1,Novello2} at $D=2$. We consider an action
$S_\Lambda$ as

\begin{equation}
S_\Lambda=\int d^2x\,\sqrt{-g}\,\,Y({\cal G})
\label{2.2}
\end{equation}

\noindent
where $g$ is the determinant of the metric tensor and $\cal G$ is
given by~\cite{Novello1}

\begin{equation}
{\cal G}=-\frac{1}{2\sqrt{-g}}\,\epsilon^{\mu\nu}F_{\mu\nu}
\label{2.3}
\end{equation}

\noindent
$F_{\mu\nu}$ is a gauge field strength (which for simplicity we are
considering to be Abelian) and $\epsilon^{\mu\nu}$ is the usual Levi-
Civita tensor density (we shall adopt the convention $\epsilon^{01}=
1$).

\medskip
The energy-momentum tensor related to the action (\ref{2.2}) is
\cite{Birrell}

\begin{eqnarray}
T^\Lambda_{\mu\nu}&=&\frac{2}{\sqrt{-g}}\,
\frac{\delta}{g^{\mu\nu}}\,\bigl(\sqrt{-g}\,Y\bigr)
\nonumber\\
&=&\frac{2}{\sqrt{-g}}\,\frac{\delta\sqrt{-g}}{\delta g^{\mu\nu}}\,Y
+2\,\frac{dY}{d{\cal G}}\,\frac{\delta{\cal G}}{\delta g^{\mu\nu}}
\nonumber\\
&=&\Bigl({\cal G}\,\frac{dY}{d{\cal G}}-Y\Bigr)\,g_{\mu\nu}
\label{2.4}
\end{eqnarray}

Considering that \cite{Novello2}

\begin{equation}
Y=\frac{k}{4}\,{\cal G}^{p+1}
\label{2.5}
\end{equation}

\noindent
where $k$ is a constant and $p$ is some rational number, we have

\begin{equation}
T^\Lambda_{\mu\nu}=\frac{kp}{4}\,{\cal G}^{p+1}g_{\mu\nu}
\label{2.6}
\end{equation}

\noindent
This expression permit us to identify the cosmological {\it constant}

\begin{equation}
\Lambda=\frac{kp}{4}\,{\cal G}^{p+1}
\label{2.7}
\end{equation}

\noindent
which can also be related to the trace of $T^\Lambda_{\mu\nu}$ as

\begin{equation}
\Lambda=g^{\mu\nu}T^\Lambda_{\mu\nu}=T^\Lambda
\label{2.8}
\end{equation}

\noindent
Notice that for $p=0$, there is no cosmological {\it constant} and for
$p=-1$, it is actually a constant.

\medskip
Introducing the cosmological constant generated in this way into the
Jackiw-Teitelboim equation, we get

\begin{equation}
R-T^\Lambda=8\pi G\tilde T
\label{2.9}
\end{equation}

\noindent
where we have absorbed a factor $8\pi G$ into the constant $k$ of the
Lagrangian density $Y$. $\tilde T$ is the trace of the remaining part
of the energy-momentum tensor of the theory (including other terms
besides $S_\Lambda$). We emphasize that in our formalism the
cosmological term was generated by the dynamics.

\medskip
In the next section we use this model to an specific example
involving a general electrodynamic theory at $D=2$.

\section{Cosmological term with electrodynamics at D=2}
\renewcommand{\theequation}{3.\arabic{equation}}
\setcounter{equation}{0}

Let us consider the general action at involving fermions and Abelian
gauge fields interacting at $D=2$,

\begin{equation}
S=\int d^2x\sqrt{-g}\Bigl[Y({\cal G})
-\frac{1}{4}F_{\mu\nu}F^{\mu\nu}
+i\bar\psi\gamma^\mu\bigl(\nabla_\mu-ieA_\mu\bigr)\psi\Bigr]
\label{3.1}
\end{equation}

\noindent
where $\nabla_\mu$ is the covariant derivative.

\medskip
Even though the Jackiw-Teiltelboim equation (\ref{2.9}) does not come
from (\ref{3.1}), we have used it as a source for the energy-momentum
tensor and, particularly, for the cosmological {\it constant}.
Considering that $T^\Lambda=\Lambda$ is given by the combination of
(\ref{2.3}) and (\ref{2.7}), and that $\tilde T=-\frac{1}{2}F_{\mu\nu}
F^{\mu\nu}$ is related to the electromagnetic energy-momentum tensor
(whose trace is not zero at $D=2$), we have for the Jackiw-Teitelboim
equation

\begin{equation}
R-\frac{kp}{4}\,
\Bigl(-\frac{1}{2\sqrt{-g}}\,\epsilon^{\mu\nu}F_{\mu\nu}\Bigr)^{p+1}
=-4\pi\,G\,F_{\mu\nu}F^{\mu\nu}
\label{3.2}
\end{equation}

Now, it is necessary to combine (\ref{3.2}) with the other equations
obtained from (\ref{3.1}), which are

\begin{eqnarray}
&&\gamma^\mu\,\biggl(\nabla_\mu-ie\,A_\mu\biggr)\,\psi=0
\label{3.3}\\
&&\sqrt{-g}\,F^{\mu\nu}_{\phantom{\mu\nu};\nu}
+\frac{k}{4}\,(p+1)\,\epsilon^{\mu\nu}\,
\Bigl[\Bigl(-\frac{1}{2\sqrt{-g}}\,
\epsilon^{\sigma\rho}F_{\sigma\rho}\Bigr)^p\Bigr]_{,\nu}
\nonumber\\
&&\phantom{\sqrt{-g}\,F^{\mu\nu}_{\phantom{\mu\nu};\nu}}
-e\,\sqrt{-g}\,\bar\psi\gamma^\mu\psi=0
\label{3.4}
\end{eqnarray}


\noindent
where the notation $_{;\mu}$ and $_{,\mu}$ means covariant and usual
derivatives respectively.

\medskip
Let us consider the metric

\begin{equation}
ds^2=dt^2-a^2(t)\,dx^2
\label{3.5}
\end{equation}

\noindent
So, the Jackiw-Teitelboim equation (\ref{3.2}) leads to

\begin{equation}
\frac{\ddot a}{a}
-\frac{kp}{4}\,\Bigl(\frac{F_{10}}{a}\Bigr)^{p+1}
=8\pi\,G\,\Bigl(\frac{F_{01}}{a}\Bigr)^2
\label{3.7}
\end{equation}

For the kind of metric given by (\ref{3.5}), we have that all fields
are just function of $t$. Thus, if one takes $\mu=0$ into (\ref{3.3}),
and considering that derivatives of fields with respect to $x$ are
zero, we get

\begin{eqnarray}
&&\psi^\dagger\psi=0
\nonumber\\
&\Rightarrow&\psi=0
\label{3.8}
\end{eqnarray}

\noindent
Now, taking $\nu=0$ into (\ref{3.3}) and using (\ref{3.8}), we also
obtain

\begin{equation}
\frac{F_{10}}{a}+\frac{k}{4}(p+1)\Bigl(\frac{F_{10}}{a}\Bigr)^p=E_0
\label{3.9}
\end{equation}

\noindent
where $E_0$ is a constant. We observe that the value of $F_{10}/a$
depends on $p$, but, for any $p$, it is always a constant.
Consequently, this model has generated an actually cosmological
constant and $\tilde T$ is also a constant. In this case, the Jackiw-
Teilteboim equation (\ref{3.7}) leads to well-known
results~\cite{Mann}.

\section{Using another Chern-Simons term}
\renewcommand{\theequation}{4.\arabic{equation}}
\setcounter{equation}{0}

The quantity we have used in the previous section is considered to be
the Chern-Simons term at $D=2$. However, it is possible to have
another sequence of Chern-Simons terms at any spacetime dimensions
\cite{Barcelos}, where for $D=2$ this term is (in curved space)

\begin{equation}
{\cal G}=-\frac{1}{2\sqrt{-g}}\,\epsilon^{\mu\nu}F_{\mu\nu}\phi
\label{4.1}
\end{equation}

\noindent
where $\phi$ is a scalar field. We observe that $T^\Lambda_{\mu\nu}$
is still given by (\ref{2.7}), but with $\cal G$ replaced by the new
one. So, the cosmological {\it constant} is

\begin{equation}
\Lambda=\frac{kp}{4}\,\Bigl(\frac{F_{10}}{a}\phi\Bigr)^{p+1}
\label{4.2}
\end{equation}

\noindent
Also here we have that for $p=0$ there is no cosmological {\it
constant} and for $p=-1$ the cosmological term is actually a constant.
The Jackiw-Teitelboim equation turns to be

\begin{equation}
\frac{\ddot a}{a}
-\frac{kp}{4}\Bigl(\frac{F_{10}}{a}\phi\Bigr)^{p+1}
=8\pi\,G\,\Bigl(\frac{F_{01}}{a}\Bigr)^2
\label{4.3}
\end{equation}

Now, the action we have to use is

\begin{equation}
S=\int d^2x\,\sqrt{-g}\,\Bigl[Y({\cal G})
-\frac{1}{4}F_{\mu\nu}F^{\mu\nu}
+\frac{1}{2}g^{\mu\nu}\phi_{,\mu}\phi_{,\nu}\Bigr]
\label{4.4}
\end{equation}

\noindent
We have not included the fermionic field because it will not
contribute according the metric convention we are using.

\medskip
As in the previous case, the Jackiw-Teitelboim equation has to be
combine with the other ones obtained from the action. These equations
are now given by

\begin{eqnarray}
&&\frac{F_{10}}{a}
+\frac{k}{4}(p+1)\phi\Bigl(\frac{F_{10}}{a}\phi\Bigr)^p=E_0
\label{4.5}\\
&&\dot a\dot\phi+a\ddot\phi
-\frac{k}{4}(p+1)F_{10}\Bigl(\frac{F_{10}}{a}\phi\Bigr)^p=0
\label{4.6}
\end{eqnarray}

The set we have to use is given by Eqs. (\ref{4.3}), (\ref{4.5}), and
(\ref{4.6}). The complete solution of this set of equations is not an
easy task. However, one can infer the asymptotic behavior of the
cosmological constant, and consequently the solution, as
$a\rightarrow\infty$ and closed to the origin.

\medskip
In the previous section, we just had solutions where $F_{10}/a$ was a
constant, independently of the region of the space. This was given by
solving Eq. (\ref{3.9}). The corresponding relation in the present
section is given by Eq. (\ref{4.5}). Now, $F_{10}/a$ is not
necessarily a constant for all values of $a$ by virtue of the presence
of $\phi$ into the second term. One observes in (\ref{4.5}) that there
is a possibility of asymptotic solution $F_{10}/a=E_0$ if
$\phi\rightarrow0$ (for $p>-1$). We also observe that this solution is
consistent with Eq. (\ref{4.6}) and leads to a nice behavior for the
cosmological {\it constant} that appear in the dynamical Jackiw-
Teitelboim equation (\ref{4.3}), where it goes to zero asymptotically.

\medskip
The behavior closed to the origin can present two typical situations:
a singularity when $a(0)=0$ or a minimum when $a(0)=a_0$. The singular
case is more restrictive. Eq. (\ref{4.5}) imposes that this situation
is only possible if $p=1$ and $k<0$. If $a(0)=a_0$ then equation
(\ref{4.5}) only determines the value of $E_0$ in terms of $F_{10}
/a_0$ and $\phi_0$. The condition that $a_0$ is a minimum impose that
$\dot a_0=0$, then Eq. (\ref{4.6}) only determines $\ddot\phi$ in
terms of $\phi_0$, $F_{10}/a_0$ and the parameters $k$ and $p$. In
addition, the condition for the existence of a minimum at $t=0$,
implies that the model exhibts an accelerated expansion. So,
considering Eq. (\ref{4.3}), we have

\begin{equation}
\frac{\ddot a_0}{a_0}=\frac{kp}{4}\,
\Bigl(\frac{F_{10}}{a_0}\,\phi_0\Bigr)^{p+1}
+8\pi\,G\,\Bigl(\frac{F_{10}}{a_0}\Bigr)^2>0.
\end{equation}

\noindent
which does not fix the sign of $\Lambda$.

\section{Conclusion}

In this paper we have studied the behavior of a time-dependent
cosmological constant in $D=2$ by using the Jackiw-Teitelboim
cosmology. We have considered two starting Lagrangians to generate the
cosmological term. The first one was based on the topological quantity
${\cal G}=\epsilon^{\mu\nu}F_{\mu\nu}/\sqrt{-g}$, which led to a
cosmological term that was actually a constant. This part was
mimicking what we have done in previous work for $D=4$. In the second
case we have considered another topological quantity, given by
${\cal G}=\phi\epsilon^{\mu\nu}F_{\mu\nu}/\sqrt{-g}$. The cosmological
term so obtained had a nice behavior, which we would like to have in a
four-dimensional spacetime. It went to zero asymptotically (with the
possibility of having a positive acceleration) and with singular or
non singular solution closed to the origin.

\medskip
This behavior we have obtained into the second case, where the
topological term involves two fields, suggests us to use a similar
procedure in $D=4$. Here, the topological would be ${\cal G}=
\epsilon^{\mu\nu\rho\lambda}F_{\mu\nu}B_{\rho\lambda}/4\sqrt{-g}$,
where $B_{\mu\nu}$ is the Kalb-Ramon field \cite{Kalb}. This work is
presently under study and possibles results shall be reported
elsewhere.

\bigskip
\begin{acknowledgments}
This work is supported in part by Conselho Nacional de Desenvolvimento
Cient\'{\i}fico e Tecnol\'ogico - CNPq (Brazilian Research Agency).
One of us, J.B.-N. has also the support of PRONEX 66.2002/1998-9.
\end{acknowledgments}

\end{document}